\documentclass{llncs}
\usepackage[utf8]{inputenc}
\usepackage[T1]{fontenc}

\usepackage{lmodern}

\usepackage{makeidx}  
\usepackage[dvipsnames]{xcolor}
\usepackage{xspace}
\usepackage{graphicx}
\usepackage{amsmath}
\usepackage{amsfonts}
\usepackage{amstext}
\usepackage{stmaryrd}
\usepackage{gensymb}
\usepackage{hyperref}
\usepackage{caption}
\usepackage[capitalise]{cleveref}

\usepackage{amsthm}

\newcommand{\slopes}[1]{{\ensuremath{Sl(#1)}}}

\newcommand{\NN}{\mathbb{N}}
\newcommand{\ZZ}{\mathbb{Z}}
\newcommand{\QQ}{\mathbb{Q}}

\newcommand{\subshift}[1]{{\ensuremath{X_{#1}}\xspace}}
\newcommand{\slopesd}{{\ensuremath{(\QQ\cup\{\infty\})^{d-1}}\xspace}}
\newcommand{\slopesdeux}{{\ensuremath{(\QQ\cup\{\infty\})^{2}}\xspace}}

\newcommand{\pizu}{{\ensuremath{\Pi_1^0}}\xspace}
\newcommand{\pizd}{{\ensuremath{\Pi_2^0}}\xspace}

\newcommand{\pizn}{{\ensuremath{\Pi_n^0}}\xspace}
\newcommand{\sizn}{{\ensuremath{\Sigma_n^0}}\xspace}
\newcommand{\sizu}{{\ensuremath{\Sigma_1^0}}\xspace}
\newcommand{\sizd}{{\ensuremath{\Sigma_2^0}}\xspace}

\makeatletter
\theoremstyle{plain}
\newtheorem*{rep@theorem}{\rep@title}
\newcommand{\newreptheorem}[2]{%
  \newenvironment{rep#1}[1]{%
     \def\rep@title{#2 \ref{##1}}%
     \begin{rep@theorem}}%
   {\end{rep@theorem}}
   }
\makeatother

\newreptheorem{theorem}{Theorem}

\theoremstyle{definition}
\newtheorem*{theoremnn}{Theorem}

\newcommand{\fig}[1]{Figure~\ref{fig:#1}\xspace}

\begin{document}
\mainmatter              
\title{Slopes of 3-dimensional Subshifts of Finite Type}
\titlerunning{Slopes of 3D subshifts}  
%
\author{Etienne Moutot\inst{1} \and Pascal Vanier\inst{2}}
\authorrunning{E. Moutot and P. Vanier} 
%
%
\institute{LIP, ENS de Lyon -- CNRS -- INRIA -- UCBL -- Universit\'e de Lyon, \\
6 all\'ee d’Italie, 69364 Lyon Cedex, France\\
\email{etienne.moutot@ens-lyon.org}
\and
Laboratoire d'Algorithmique, Complexit\'e et Logique\\
Universit\'e de Paris-Est, LACL, UPEC, France\\
\email{pascal.vanier@lacl.fr}}

\maketitle              

\begin{abstract}
    In this paper we study the directions of periodicity of three-dimensional subshifts of
    finite type (SFTs) and in particular their slopes. A configuration of a subshift has a slope of
    periodicity if it is periodic in exactly one direction, the slope being the angles of the
    periodicity vector. In this paper, we prove that any $\Sigma^0_2$ set may be realized as a
    a set of slopes of an SFT.
\end{abstract}

A $d$-dimensional subshift of finite type (SFT for short) is a set of colorings of $\ZZ^d$ by a
finite number of colors containing no pattern from a finite family of forbidden patterns.
Subshifts may be seen as discretizations of continuous dynamical systems: if $X$ is a compact
space and there are $d$ commuting continuous actions $\phi_1,\dots, \phi_d$ on $X$, one can
partition $X$ in a finite number of parts indexed by an alphabet $\Sigma$. The orbit of
a point $x\in X$ maps to a coloring $y$ of $\ZZ^d$ where $y(v)$ corresponds to the
partition where $\phi^v(x)$ lies.

In dimension 1, most problems on SFTs are easy in a computational sense, since SFTs correspond
to bi-infinite walks on finite automata. For instance, in dimension 1, detecting whether an SFT is
non-empty is decidable since it suffices to detect if there exists a cycle in the corresponding
automaton \cite{LindMarcus}, which corresponds to the existence of a periodic configuration.

In higher dimensions however, the situation becomes more involved, and knowing whether an SFT is
non-empty becomes undecidable \cite{Berger2,BergerPhD}. The proof uses two key results on SFTs: the existence of an aperiodic SFT and an encoding of Turing machine's space time diagrams. The
fact that there exists aperiodic SFTs is not straightforward, and the converse was first
conjectured by Wang \cite{WangI}. Had this conjecture been true, it would have meant the
decidability of the emptiness problem for SFTs. Berger \cite{Berger2,BergerPhD} proved however that there
does exist SFTs containing only aperiodic configurations. Subsequently, many other aperiodic
SFTs were constructed \cite{Robinson,Kari14,Ollinger,Poupet,wang11,CulikKariCubes}. Note that the existence
in itself of aperiodic SFTs does not suffice to prove that the emptiness problem is undecidable,
one needs in addition to encode some computation in them, usually in the form of Turing machines.

Periodicity has thus been central in the study of SFTs from the beginning, and it has
been proved very early that knowing whether an SFT is aperiodic is undecidable \cite{Gure}.
In fact, sets of periods
constitute a classical conjugacy/isomorphism invariant for
subshifts in any dimension. As such, they have been
studied extensively and even characterized: algebraically in dimension 1, see \cite{LindMarcus} for more details, and computationally in dimension 2. In fact it seems that computability theory is
the right tool to study dynamical aspects of higher dimensional symbolic dynamical systems \cite{HochMey,Meyero,Aubrun2,DuRoSh}.

In dimensions $d\geq 2$, one may investigate periodicity from different angles.
Denote  $\Gamma_x = \{v\in \ZZ^d\mid x(z+v)=x(z), \forall z\in\ZZ^d\}$ the lattice of vectors of periodicity
of configuration $x$: $\Gamma_x$ may be of any dimension below $d$ and some cases are particularly
interesting:
\vspace{-1em}
\begin{itemize}
    \item When it is of dimension 0, then $x$ does not have any vector of periodicity and is hence
        aperiodic.
    \item When it is of dimension $d$, then $x$ is somehow finite, this case has been studied
        and partly characterized in terms of complexity classes by Jeandel and Vanier \cite{JeandelV2013b}.
    \item When $d=1$, then there exists some vector $v$ such that $\Gamma_x=v\ZZ$. In this case,
        one may talk about the direction or slope of the configuration.
\end{itemize}

In this paper, we are interested in this last case. In \cite{JeandelV2010a}, this case was studied
 and characterized for 2-dimensional SFTs through the arithmetical hierarchy:
 \begin{theoremnn}[Jeandel and Vanier]
     The sets of slopes of 2-dimensional SFTs are exactly the \sizu subsets of $\QQ\cup\{\infty\}$.
 \end{theoremnn}

 In the end of \cite{JeandelV2010a} it was conjectured that slopes of higher dimensional SFTs are
 the \sizd subsets of \slopesd. This gap between dimension 2 and dimension 3 for decidability of
 periodicity questions is similar to the gap
 between dimension 1 and 2 for decidability of emptiness questions: the subset of periodic
 configurations of a $d$-dimensional subshift along some periodicity vector may be seen as
 a $(d-1)$-dimensional subshift (see \emph{e.g.} \cite{JeandelV2010a}), hence the jump in
 complexity. This is the idea that led to the conjecture. However, in dimension higher than 2, the construction of \cite{JeandelV2010a} cannot be reused.

In this article, we prove one direction of the aforementioned conjecture: we show how to realize any \sizd subset of \slopesdeux~as a set of slopes of a 3D subshift:
\begin{theorem}
  \label{th:sigma2}
  Any \sizd subset of \slopesdeux~may be realized as the set of slopes of some 3D SFT.
\end{theorem}
In order to do this, we introduce a new way to synchronize computations between different
dimensions, inspired partly by what is done by Durand, Romashchenko and Shen \cite{DuRoSh}.
Note that our construction can be easily generalized to realize any \sizd subset of \slopesd~as a set of slopes of a $d$-dimensional subshift for $d\geq 3$.

However, we did not manage to prove the other part of the conjecture, that is the fact that the sets of slopes of $d$-dimensional SFTs are in \sizd (for $d\geq 3$).

The paper is organized as follows: in section~\ref{def} we recall the useful definitions about
subshifts and the arithmetical hierarchy, and
in section~\ref{S:sigma2} we prove Theorem~\ref{th:sigma2}.

\section{Definitions and properties}\label{def}
\subsection{Subshifts and tilesets}
We give here some standard definitions and facts about subshifts, one may consult \cite{LindMarcus}
for more details.

 Let $\Sigma$ be a finite alphabet, a \emph{configuration} (or tiling) is a function
 $c:\ZZ^d\longrightarrow \Sigma$. A \emph{pattern} is a function
 $p:N \longrightarrow \Sigma$, where $N\subseteq \ZZ^d$ is a finite set,
  called the \emph{support} of $p$. A pattern $p$ \emph{appears} in another pattern $p'$ if there exists $v\in \ZZ^d$ such that ${\forall x\in N}, {p(x)=p'(x+v)}$.
  We write then $p\sqsubseteq p'$.
Informally, a configuration (or tiling) is a coloring of $\ZZ^d$ with elements of $\Sigma$.
A \emph{subshift} is a closed, shift-invariant subset of $\Sigma^{\ZZ^d}$, the $d$-dimensional
\emph{full shift}. For a subshift $X$ we will sometimes note $\Sigma_X$ its alphabet.
The full shift is a compact metric space when equipped with the distance
$d(x,y) = 2^{-\min\{\|v\|_\infty ~|~ v\in\ZZ^d, x(v)\neq y(v)\} }$
with $\|v\|_\infty = \max_i |v_i|$.

It is well known that subshifts may also be defined via collections of
forbidden patterns. Let $F$  be a collection of forbidden patterns, the subset $\subshift{F}$ of
$\Sigma^{\ZZ^d}$ defined by
\[
    \subshift{F}=\left\{ x\in\Sigma^{\ZZ^d}\mid \forall p\in F, p\not\sqsubseteq x\right\}
\]
is a subshift. Any subshift may be defined via an adequate collection of forbidden patterns.
A \emph{subshift of finite type (SFT)} is a subshift which may be defined via a finite collection
of forbidden patterns. A configuration of a subshift is also called a \emph{point} of this subshift
and is said to be valid with respect to the family of forbidden patterns $F$.
Remark that $F$ being finite, we can define a subshift of finite type either by a set of forbidden or authorized patterns.

\emph{Wang tiles} are unit squares with colored edges which may not be flipped or rotated, a
\emph{tileset} is a finite set of Wang tiles. Tiles of a tileset maybe placed side by side on the
$\ZZ^2$ plane only when the matching borders have the same color, thus forming a tiling of the plane.
The set of all tilings by some tileset is an SFT, and conversely, any SFT may be converted into an
isomorphic tileset. From a computability point of view, both models are equivalent and we will use
both indiscriminately. In 3D, Wang tiles can be straightforwardly generalized to Wang cubes.

A subshift is \textbf{North-West-deterministic} if, for any position, and for any two colors placed
above it and to its left, there exists at most one valid color at this position.
  Likewise, we call a subshift \textbf{West-deterministic} if it is the case with the colors to its
  left and top-left.

  \begin{figure}[h]
  \centering
  \begin{minipage}{.45\textwidth}
    \centering
    \includegraphics{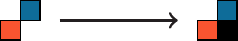}
    \captionof{figure}{NW-determinism.}
    \label{fig:def:NWdet}
  \end{minipage}%
  \begin{minipage}{.1\textwidth}
      ~
  \end{minipage}%
  \begin{minipage}{.45\textwidth}
    \centering
    \includegraphics{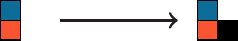}
    \captionof{figure}{W-determinism.}
    \label{fig:def:Wdet}
  \end{minipage}
  \end{figure}

\subsection{Periodicity and aperiodicity}
The notion of periodicity being central in this paper, we will define it in this section.

\begin{definition}[Periodicity]
    A configuration $c$ is \emph{periodic} of period $v$ if there exists $v\in\ZZ^d\setminus\{(0,0)\}$ such that $\forall x\in\ZZ^d, c(x) = c(x+v)$. If $c$ has no period, then it is said to be
    \emph{aperiodic}. A subshift is \emph{aperiodic} if all its points are aperiodic.
\end{definition}

From now on, we will focus on dimension 3 in this paper.
As seen in the introduction, the lattice of vectors of periodicity may be of any dimension between
0 and $d$ and we are interested here in the case where it is 1-dimensional. In this case we
can define the slope periodicity:

\begin{definition}[Slope of periodicity]
  Let $c$ be a configuration periodic along $v=(p,q,r)$.
  We call \emph{slope of $v$} the pair $\theta = (\theta_1, \theta_2)$ with $\theta_1=\frac{p}{r}$ and $\theta_2=\frac{p}{q}$.
  If all vectors of periodicity of $c$ have slope $\theta$, we say that $\theta$ is the \emph{slope of periodicity} or \emph{slope} of $c$.
  We write $\slopes{X} = \left\{\theta \mid \exists x\in X, \theta \text{ is the slope of }x\right\}$ the set of slopes of $X$.
\end{definition}

\subsection{Arithmetical hierarchy}
We give now some basic definitions used in computability theory and in particular about the arithmetical
hierarchy. More details may be found in \cite{Rogers}.

Usually the arithmetical hierarchy is seen as a classification of sets according to their logical characterization.
For our purpose we use an equivalent definition in terms of computability classes and Turing machines with oracles:

  \begin{itemize}
  \item $\Delta^0_0 = \Sigma_0^0 = \Pi_0^0$ is the class of recursive (or computable) problems.
    \item \sizn is the class of recursively enumerable (RE) problems with an oracle $\Pi_{n-1}^0$.
    \item \pizn the complementary of \sizn, or the class of co-recursively enumerable (coRE) problems with an oracle $\Sigma_{n-1}^0$.
    \item $\Delta^0_n = \sizn \cap \pizn$ is the class of recursive (R) problems with an oracle $\Pi_{n-1}^0$.
  \end{itemize}

In particular, \sizu is the class of recursively enumerable problems and \pizu is the class of
co-recursively enumerable problems.

\section{Proof of \cref{th:sigma2}}\label{S:sigma2}

\begin{reptheorem}{th:sigma2}
    Let $R \in \sizd\cap \mathcal{P}((\QQ\cup \{\infty\})^2)$, there exists a 3D SFT $X$ such that $\slopes{X} = R$.
\end{reptheorem}

\begin{proof}
Let $M$ be a Turing machine accepting $R$ with an oracle $O\in \pizu$. One can suppose that this machine takes as input 3 integers $(p,q,r)\in\NN^3$ and that its output depends only on $\theta_1=\frac{p}{r}$ and $\theta_2=\frac{p}{q}$.

We only explain the case $0<r<q<p$, the others are symmetric or quite similar and it suffices to take the
disjoint union of the obtained SFTs to get the full characterization.

Let us construct a 3D SFT $X_M$ that has a periodic configuration along $\theta$ if and only if $\theta\in R$.
To do so, $X_M$ will be such that "good" configurations (i.e. valid and 1-periodic) are formed of large cubes, shifted with an offset to allow periodicity along some slope.
Then we encode $M$ inside all the cubes, and give to it the slope as input. The machine halts (i.e the slope is in $R$) implies that the cubes are of finite size. Which means that the configuration is 1-periodic only when the slope actually corresponds to some element of $R$.

For that, we separate the construction in different layers, in order to make it clearer.
We define $X_M=B \times B' \times B'' \times C \times W \times P \times S \times T_O \times T_M \times A$, with the following layers:
\begin{itemize}
  \item $B$ creates $(yz)$ black planes, separated by an aperiodic tiling.
  \item $B'$ and $B''$ create planes orthogonal to the ones of $B$, forming rectangular parallelepipeds.
  \item $C$ forces the parallelepipeds to become cubes.
  \item $W$ forces the aperiodicity vector to appear between cubes, and writes the input of the Turing
      machine in the cubes.
  \item $P$ reduces the size of the input.
  \item $S$ synchronizes the aperiodic backgrounds of the cubes.
  \item $T_M$ encodes the "\sizd" Turing machine $M$ in the cubes.
  \item $T_O$ encodes the \pizu  oracle $O$ that is used by $M$.
  \item $A$ ensures the existence of configurations with a unique direction of periodicity.
\end{itemize}

\begin{description}
  \item[Aperiodic background] We first need an aperiodic background in order to ensure that there is no other directions of periodicity that the one we create later on. We even make a 3D West-deterministic aperiodic background since some layers will
      need that deterministic property to work. For that we cross two 2D West-deterministic aperiodic
      backgrounds: the set of aperiodic cubes are the sets of cubes of the form shown in \fig{def:aperiodique3D}.
      We also impose that all parallel planes are identical (\fig{def:aperiodique3D_copie}). The 2D aperiodic
      tiling is from Kari \cite{kariNW}, which is a NW-aperiodic tiling, and can be easily transformed into
      a West-deterministic SFT. With such a superposition, one can easily show that the resulting 3D tiling is aperiodic.

    \begin{figure}[hb]
    \centering
    \begin{minipage}{.45\textwidth}
      \centering
      \includegraphics{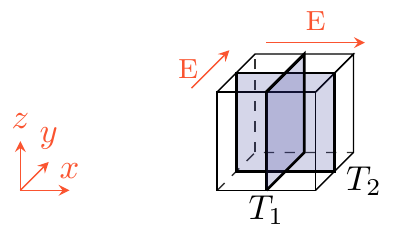}
      \captionof{figure}{$T_1$, $T_2$ are tiles of the 2D aperiodic tileset and form a Wang cube of the 3D
          aperiodic tileset. The "E" shows the east direction of the two planes.}
      \label{fig:def:aperiodique3D}
    \end{minipage}%
    \begin{minipage}{.1\textwidth}
        ~
    \end{minipage}%
    \begin{minipage}{.45\textwidth}
      \centering
      \includegraphics{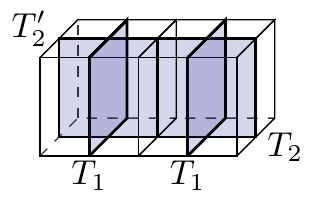}
      \captionof{figure}{Duplication of parallel backgrounds.}
      \label{fig:def:aperiodique3D_copie}
    \end{minipage}
    \end{figure}

    \item[Layer $B$] The first layer is made with two types of cubes: a white cube
        (\includegraphics[height=9pt]{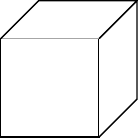}), which is a meta-cube that
        corresponds to any cube of the aperiodic background and a black cube
        (\includegraphics[height=9pt]{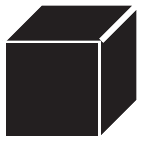}) which will serve to break the
        aperiodicity brought by the white cubes.
        The rules of this layer are:
    \begin{itemize}
      \item In coordinates $z+1$ and $z-1$ of
          \includegraphics[height=9pt]{figures/pentes_1/tuiles/noir.pdf}, only a
          \includegraphics[height=9pt]{figures/pentes_1/tuiles/noir.pdf} can appear.
      \item In coordinates $y+1$ and $y-1$ of
          \includegraphics[height=9pt]{figures/pentes_1/tuiles/noir.pdf}, only
          \includegraphics[height=9pt]{figures/pentes_1/tuiles/noir.pdf} can appear.
      \item In coordinates $x+1$ and $x-1$ of
          \includegraphics[height=9pt]{figures/pentes_1/tuiles/noir.pdf}, only
          \includegraphics[height=9pt]{figures/pentes_1/tuiles/blanc.pdf} can appear.
    \end{itemize}
    With only this layer, the valid periodic configurations are thick aperiodic $(yz)$ planes separated by
    infinite black $(yz)$ planes. At this stage, there may be several aperiodic planes "inside" a period.

    \item[Layer $B'$]
    For every Wang cube of this layer, we impose that the cube at $y+1$ is the same.
    so we can describe the layer $B'$ by a set of 2D Wang tiles in the $(xz)$ plane, duplicated on the $y$ axis.
    The tiles are:
    \begin{center}
      \includegraphics[width=20pt]{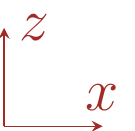}
      ~~~~~
      \includegraphics[width=10pt]{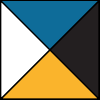}
      \includegraphics[width=10pt]{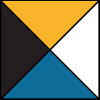}
      \includegraphics[width=10pt]{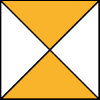}
      \includegraphics[width=10pt]{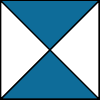}
      ~~~~~
      \includegraphics[width=10pt]{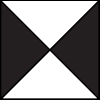}
      \includegraphics[width=10pt]{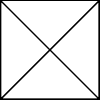}
      ~~~~~~~~~
    \end{center}
    The first four can only be superimposed with black tiles of layer $B$ and the last two only with white ones.

    With layers $B$ and $B'$ the periodic configurations are formed of infinite planes along $(yz)$ linked by
    infinite $(xy)$ strips infinite along $y$, see Figure~\ref{fig:pentes_1:stepBp}.
    \begin{figure}[h]
    \begin{minipage}{.45\textwidth}
      \begin{center}
        \includegraphics{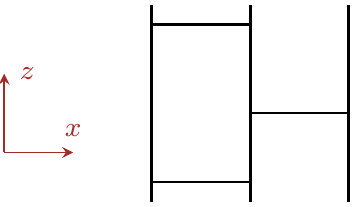}
        \caption{Projection on the $(xz)$ plane of a valid configuration with layers $B$ and $B'$.}
        \label{fig:pentes_1:stepBp}
      \end{center}
    \end{minipage}
    \begin{minipage}{.1\textwidth}
        ~
    \end{minipage}
    \begin{minipage}{.45\textwidth}
      \begin{center}
        \includegraphics[scale=0.4]{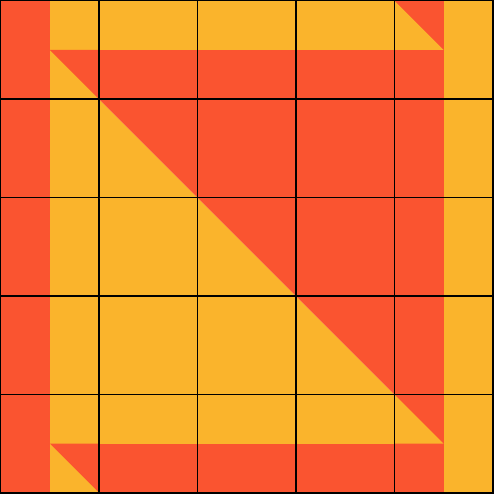}
        \caption{Valid tiling with rules of layer $C$.}
        \label{fig:pentes_1:stepC}
      \end{center}
    \end{minipage}
    \end{figure}

    \item[Layer $B''$] This layer is identical to $B'$ but tiles are duplicated along the $z$ axis.
    It creates portions of infinite planes along the $z$ axis, also delimited by the
    black planes the layer $B$.

    With these three layers, periodic configurations are formed of parallelepipeds delimited by black cubes
    and \includegraphics[width=10pt]{figures/pentes_1/tuiles/B5.pdf} with portions of aperiodic background inside them.
    \item[Layer $C$] This layer forces the parallelepipeds to be cubes, by forcing rectangles of $(xz)$ to
        be squares, and same for rectangles of $(xy)$.

    Like the $B$ layer this layer is created by duplicating 2D Wang tiles along $y$ axis for rectangles of $(xz)$
    and $z$ axis for rectangles of $(xy)$:
    \begin{center}
     \includegraphics[width=10pt]{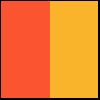}
     \includegraphics[width=10pt]{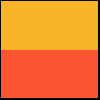}
     \includegraphics[width=10pt]{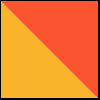}
     \includegraphics[width=10pt]{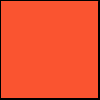}
     \includegraphics[width=10pt]{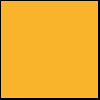}
     \includegraphics[width=10pt]{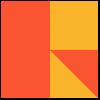}
     \includegraphics[width=10pt]{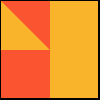}
    \end{center}
    These tiles are superimposed once on the $B'$ tiles with rules on the $(xz)$ plane and on the $B''$ tiles
    with rules on the $(xy)$ plane. The superimpositions allowed are the following:
    \begin{itemize}
      \item \includegraphics[width=7pt]{figures/pentes_1/tuiles/C1.pdf} can only be on \includegraphics[width=7pt]{figures/pentes_1/tuiles/B3.pdf} and \includegraphics[width=7pt]{figures/pentes_1/tuiles/B4.pdf}.
      \item \includegraphics[width=7pt]{figures/pentes_1/tuiles/C2.pdf} can only be on \includegraphics[width=7pt]{figures/pentes_1/tuiles/B5.pdf}
      \item \includegraphics[width=7pt]{figures/pentes_1/tuiles/C6.pdf} can only be on \includegraphics[width=7pt]{figures/pentes_1/tuiles/B1.pdf} and \includegraphics[width=7pt]{figures/pentes_1/tuiles/C7.pdf} only on \includegraphics[width=7pt]{figures/pentes_1/tuiles/B2.pdf}
      \item \includegraphics[width=7pt]{figures/pentes_1/tuiles/C3.pdf},
     \includegraphics[width=7pt]{figures/pentes_1/tuiles/C4.pdf} and
     \includegraphics[width=7pt]{figures/pentes_1/tuiles/C5.pdf} can only be on white tiles \includegraphics[width=7pt]{figures/pentes_1/tuiles/B6.pdf}
    \end{itemize}

    Figure \ref{fig:pentes_1:stepC} shows how this layer forces squares to appear.

  \item[Layer $W$]
  This layer uses signals to synchronize the offsets of different cubes, and to force cubes to have the same size.
  In order to visualize the different offsets you can refer to Figure \ref{fig:pentes_1:offsets}.
  The construction is done in several parts.

  The first one forces the offsets along $y$ (denoted by $r$) to be the same in each cube. Here again, everything is duplicated along $z$. It creates signals (see \fig{pentes_1:signals_a}), that have to correspond with the extension of the neighboring cubes. This also writes the number $r$ in unary in the border of each cube. This number will be used by the Turing machine encoded later in the tiling.

  The second part is identical to the first one, but on the $(xz)$ plane and rotated 90\degree. It forces the offset along $z$ (denoted by $q$) to be the same everywhere.

  Finally, the cubes are forced to be of same size. For that we add the two signals shown on
  \fig{pentes_1:signals_b}, which have to link a corner to the extension of a square. It has the
  effect to force each square (and hence each cube) to be of same size as its neighbors.

  \begin{figure}[h]
    \centering
    \begin{minipage}{.25\textwidth}
      \includegraphics[scale=1.2]{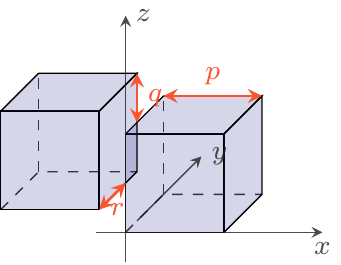}
      \captionof{figure}{Names of the offsets.}
      \label{fig:pentes_1:offsets}
    \end{minipage}
    \begin{minipage}{.1\textwidth}
        ~
    \end{minipage}%
    \begin{minipage}{.30\textwidth}
      \includegraphics[scale=0.8]{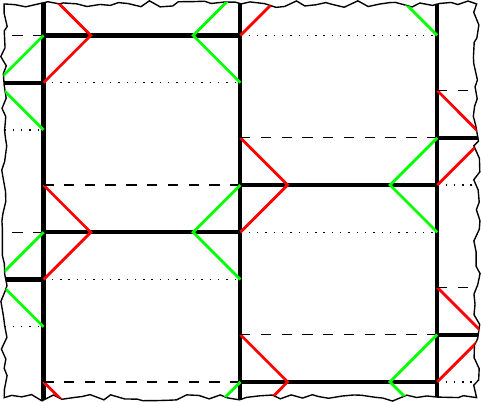}
      \captionof{figure}{Signals making the offsets identical.}
      \label{fig:pentes_1:signals_a}
      \end{minipage}%
    \begin{minipage}{.1\textwidth}
        ~
    \end{minipage}%
    \begin{minipage}{.20\textwidth}
      \includegraphics[scale=0.8]{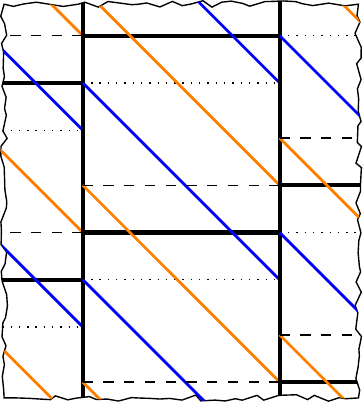}
      \captionof{figure}{Signals making the cubes of same size.}
      \label{fig:pentes_1:signals_b}
    \end{minipage}
\end{figure}

  \item[Layer $P$]
  This layer reduces the size of the input, in order to allow us to construct valid configurations as large as we
  want for the same input $(p', q', r')$.
  Starting from an unary input $(p,q,r)$, this layer writes into cubes what the input of the Turing
  machine will be: $(p', q', r')$, with $(p,q,r) = 2^k (p',q',r')$, and $\text{gcd}(p',q',r')$
  not divisible by 2.

  For that, we use a transducer to convert numbers in binary. Such a transducer can be easily encoded
  into tilings (see for example \cite{Kari14} or \cite{wang11}). Then it only remains to remove the final 0's they have in common, which can easily
  be done through local rules.

  \item[Layer $S$]\label{layerS}
  Aperiodic backgrounds of different "slices" may be different ("slices" are the thick planes in the $(yz)$ plane). They must be synchronized in order to ensure the existence of a periodic configuration along $(p,q,r)$. To do this synchronization in 2D, we use the following arrow tiles:
  \begin{center}
  \includegraphics[width=10pt]{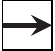}
  \includegraphics[width=10pt]{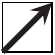}
  \includegraphics[width=10pt]{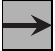}
  \includegraphics[width=10pt]{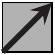}
  \end{center}
  with the following rules:
  \begin{itemize}
    \item To the left of \includegraphics[width=7pt]{figures/pentes_1/tuiles/B1.pdf} (layer B') there is \includegraphics[width=7pt]{figures/pentes_1/tuiles/haut2.pdf} and bottom left neighbor of \includegraphics[width=7pt]{figures/pentes_1/tuiles/haut2.pdf} is \includegraphics[width=7pt]{figures/pentes_1/tuiles/haut2.pdf} or a  \includegraphics[width=7pt]{figures/pentes_1/tuiles/droite2.pdf}.
    \item Bottom left tile of a square is \includegraphics[width=7pt]{figures/pentes_1/tuiles/droite2.pdf}. On the right of \includegraphics[width=7pt]{figures/pentes_1/tuiles/droite2.pdf} there is only \includegraphics[width=7pt]{figures/pentes_1/tuiles/droite2.pdf} or \includegraphics[width=7pt]{figures/pentes_1/tuiles/haut1.pdf}.
    \item On the right, left and bottom of \includegraphics[width=7pt]{figures/pentes_1/tuiles/haut1.pdf} there is only \includegraphics[width=7pt]{figures/pentes_1/tuiles/haut1.pdf} or \includegraphics[width=7pt]{figures/pentes_1/tuiles/haut2.pdf}.
    \item The \textit{breaking lines} can only have \includegraphics[width=7pt]{figures/pentes_1/tuiles/droite1.pdf} on them.
  \end{itemize}
  We obtain the tiling shown on Figure~\ref{fig:pentes_1:synchro2D}. If we impose that the background is the
  same at the beginning and at the end of the arrow with gray background, its West-periodicity ensures that it is repeated
  along the global periodicity vector.

  \vspace{-1em}

  \begin{figure}[h]
  \centering
  \begin{minipage}{.5\textwidth}
    \begin{center}
      \includegraphics[scale=0.3]{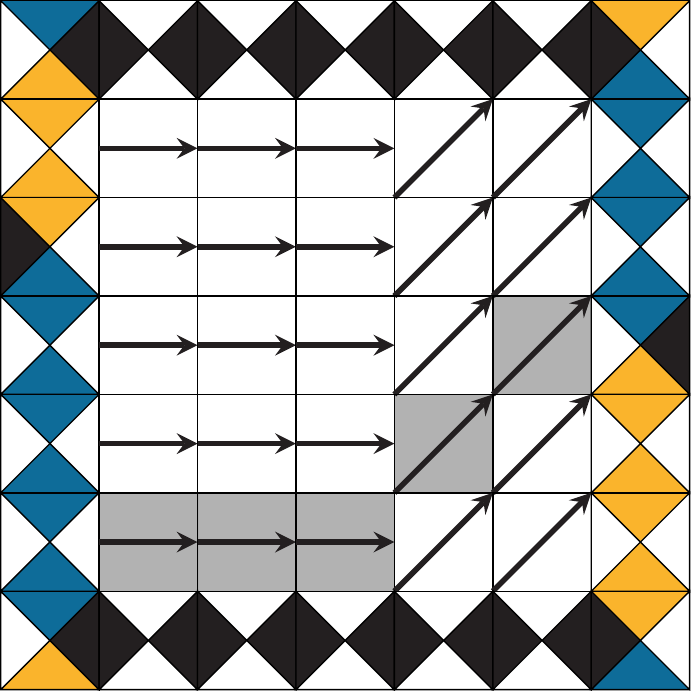}
      \caption{Transmission of a 2D background.}
      \label{fig:pentes_1:synchro2D}
    \end{center}
  \end{minipage}
  \begin{minipage}{.1\textwidth}
      ~
  \end{minipage}%
  \begin{minipage}{.38\textwidth}
    \begin{center}
      \begin{tabular}{|c|c|c|}
      \hline
        front & top & 3D \\
        \hline
         \includegraphics[width=10pt]{figures/pentes_1/tuiles/droite1.pdf} & \includegraphics[width=10pt]{figures/pentes_1/tuiles/droite1.pdf} & \includegraphics[width=12pt]{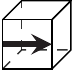}\\
          \includegraphics[width=10pt]{figures/pentes_1/tuiles/haut1.pdf} & \includegraphics[width=10pt]{figures/pentes_1/tuiles/droite1.pdf} & \includegraphics[width=12pt]{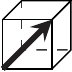}\\
          \includegraphics[width=10pt]{figures/pentes_1/tuiles/droite1.pdf} & \includegraphics[width=10pt]{figures/pentes_1/tuiles/haut1.pdf} & \includegraphics[width=12pt]{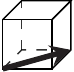}\\
          \includegraphics[width=10pt]{figures/pentes_1/tuiles/haut1.pdf} & \includegraphics[width=10pt]{figures/pentes_1/tuiles/haut1.pdf} & \includegraphics[width=12pt]{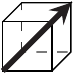}\\
         \hline
      \end{tabular}
      \caption{Rules for layer S (transmission in 3D).}
      \label{fig:pentes_1:3Dsynchro}
    \end{center}
  \end{minipage}%
  \end{figure}

  \vspace{-2em}

  We now use this 2D construction to build the 3D transmission of the background in $X_M$.
  We create two layers of 2D arrows. One in the $(xy)$ plane, that are repeated along $z$
  (\emph{front} arrows) and the other in $(xz)$ and repeated along $y$ (\emph{top} arrow).
  We then create our real layer using these two 2D layers, with 3D arrows:
  \begin{center}
  \includegraphics[width=16pt]{figures/pentes_1/tuiles/3Ddroite.pdf}
  \includegraphics[width=16pt]{figures/pentes_1/tuiles/3Dhaut.pdf}
  \includegraphics[width=16pt]{figures/pentes_1/tuiles/3Ddroitehaut.pdf}
  \includegraphics[width=16pt]{figures/pentes_1/tuiles/3Ddroitedroite.pdf}
  \end{center}
  The superimposition of two 2D arrows gives directly which 3D arrow is on each tile (see Figure~\ref{fig:pentes_1:3Dsynchro}).
  Like in 2D we impose that the background is the same at the beginning and at the end of the gray arrow. Thanks to the double West-periodicity of the background, this ensures that the background has a periodicity vector of $(p,q,r)$ in valid configurations.

  \item[Layer $T_M$] This layer encodes the Turing machine $M$ in the tiling. In our definition of the arithmetical hierarchy, the machine $M$ being in \sizd, it has access to a \pizu oracle. This oracle will be represented by a tape $R_O$ filled with zeros and ones, such that position $i$ of $R_O$ is a $1$ if and only if the oracle $O$ accepts $i$ (i.e. the Turing machine of index $i$ runs indefinitely).
  For the moment, we will encode $M$ with an additional read-only arbitrary tape, and the next layer will ensure that the content of the tape is valid for $O$.
  This additional tape is supposed to be infinite, but since $M$ has to halt in the periodic configurations, we can restrict the construction to a finite but arbitrarily large portion of it. The tape is a line along axis $y$ duplicated along axes $x$ and $z$ (see Figure~\ref{fig:pentes_1:ruban_oracle}).
  We add the two rules:
  \begin{enumerate}
    \item Inside a cube, a number at position $x$ is the same as the number at position $x-1$ and a number at position
        $z$ must be equal to the number at position $z-1$. \label{i:dupli1}
    \item The first line of the tape is transmitted through black cubes like the aperiodic background.
  \end{enumerate}
  The first rule duplicates the first line everywhere inside a cube, and the second one ensures that the same $R_O$ tape is duplicated along the direction of periodicity.

  Then, we encode $M$ in the $(xz)$ plane using the usual encoding of Turing machines in tilings.
  Let us say that the time is along the $z$ axis and the working tape along $x$.
  In order to access the entire $R_O$ tape, we add the spacial dimension $y$ to the TM encoding: while doing a transition, the machine can move its head along the $y$ axis and read the value of the $R_O$ tape in it; rule \ref{i:dupli1} above prevents $M$ to modify this extra tape.

  Note that because $O$ is a \pizu oracle, it can only ensure that the $1$s of $R_O$ are correct. $M$ has to check that the $0$s are correct. But checking the 0s, i.e. checking if a TM halts in a computation does not add any complexity to the problem, because we are only interested in the periodic configurations, where $M$ actually halts (and so do all its checks).

  \item[Layer $T_O$]

      This layer is the core of this proof, and it is where 3D actually comes to play: in the thick aperiodic
      $(yz)$ planes we will compute the \pizu oracle by encoding an infinite computation that checks
      simultaneously all possible inputs of $M_O$, the Turing machine checking the \pizu oracle $O$ ($M_O$ halts if and only if there is a wrong 1 in the portion of $R_O$ written in all the cubes).

    The key idea of this layer is the use of the previously constructed cubes as \emph{macro-tiles}
    in order to encode computations of $M_O$. Each cube will thus represent one tile and the thick aperiodic
    planes will contain, more sparsely, another 2D tiling. See Figure~\ref{fig:pentes_1:transitions_norm} to see
    how the cubes store this macro-tileset.
    For this macro-tileset, we may use a construction of
    Myers \cite{Myers} which modifies Robinson's aperiodic tileset in order to synchronize the input tapes on all of
    the partial computations. So each of our cubes contains/represents one tile of Myer's tileset, and the
    thick aperiodic planes thus also contain a Myers tiling checking some input that for the moment is
    not synchronized with the oracle written inside these cubes.

    We now have a valid macro-tiling for the cubes if and only if the machine $M_O$ never halts on $R_O$.

    The one remaining thing to do is to explain the $R_O$ tape that $M_O$ accesses is synchronized
    with the $R_O$ which is stored inside the large cubes.
    We add to the set of numbered tiles the same tiles, but in red, representing the head of
    the $M_O$ Turing machine on the tape $R_O$. We impose that there is only one red number in every large
    cube (see Figure~\ref{fig:pentes_1:marqueur_lecture}).

    The red tile of a cube must be synchronized with the cell of the oracle $R_O$ currently contained in
    the Myers tile. Every time a new partial computation is started in the macro-tiles, the red tile
    must be placed at the beginning of $R_O$, whenever the macro-tile moves the head to the right/left,
    the red tile must also be moved, if the red tile reaches the border of the cube, in which case it reaches
    a special state of non-synchronization, since the beginning has already been synchronized.

    To do that, we must allow two new transitions. These new transitions do not change the state of
    the working tape, thus we only move to the next time along $z$. But in the new position, the red-marked cell in the large cube must have changed. To do this, we again use signals between the bottom-cube (previous state), the cell doing the transition and the upper-cube (next state), see \fig{pentes_1:marqueur_lecture_move}.

    \begin{figure}[h]
    \centering
    \begin{minipage}{.45\textwidth}
      \begin{center}
        \includegraphics[scale=0.5]{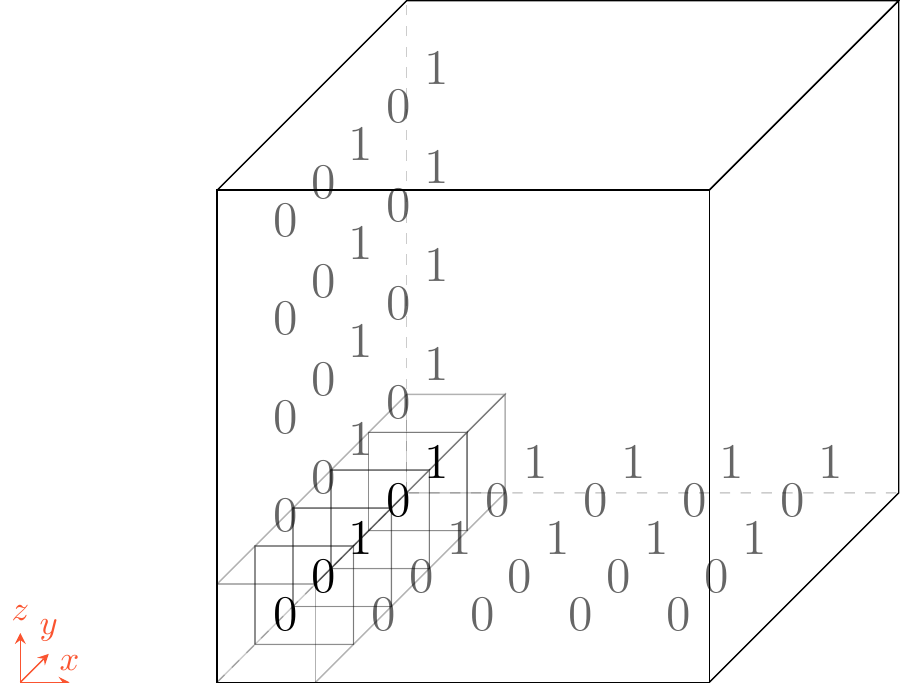}
        \caption{Tape $R_O$ of the oracle in a cube.}
        \label{fig:pentes_1:ruban_oracle}
      \end{center}
    \end{minipage}%
    \begin{minipage}{.1\textwidth}
        ~
    \end{minipage}%
    \begin{minipage}{.45\textwidth}
      \begin{center}
        \includegraphics[scale=0.6]{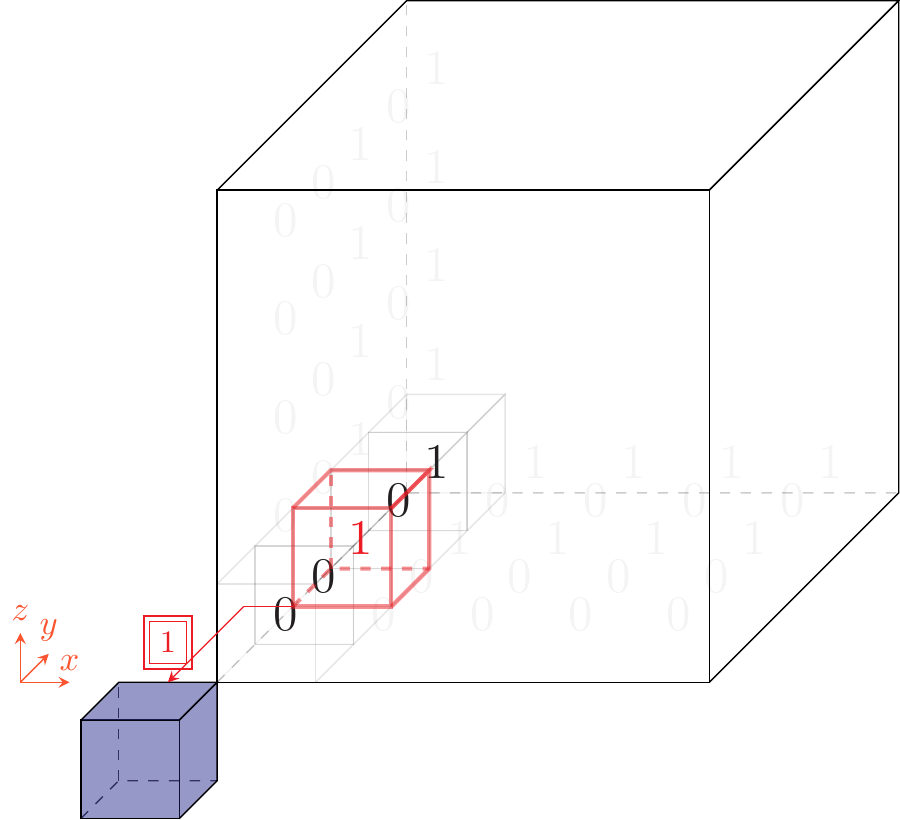}~~~
        \caption{The red tiles on $R_O$ and the transmission of its value.}
        \label{fig:pentes_1:marqueur_lecture}
      \end{center}
    \end{minipage}%
    \end{figure}

    \begin{figure}[h]
    \centering
    \begin{minipage}{.44\textwidth}
      \begin{center}
        \includegraphics[scale=0.48]{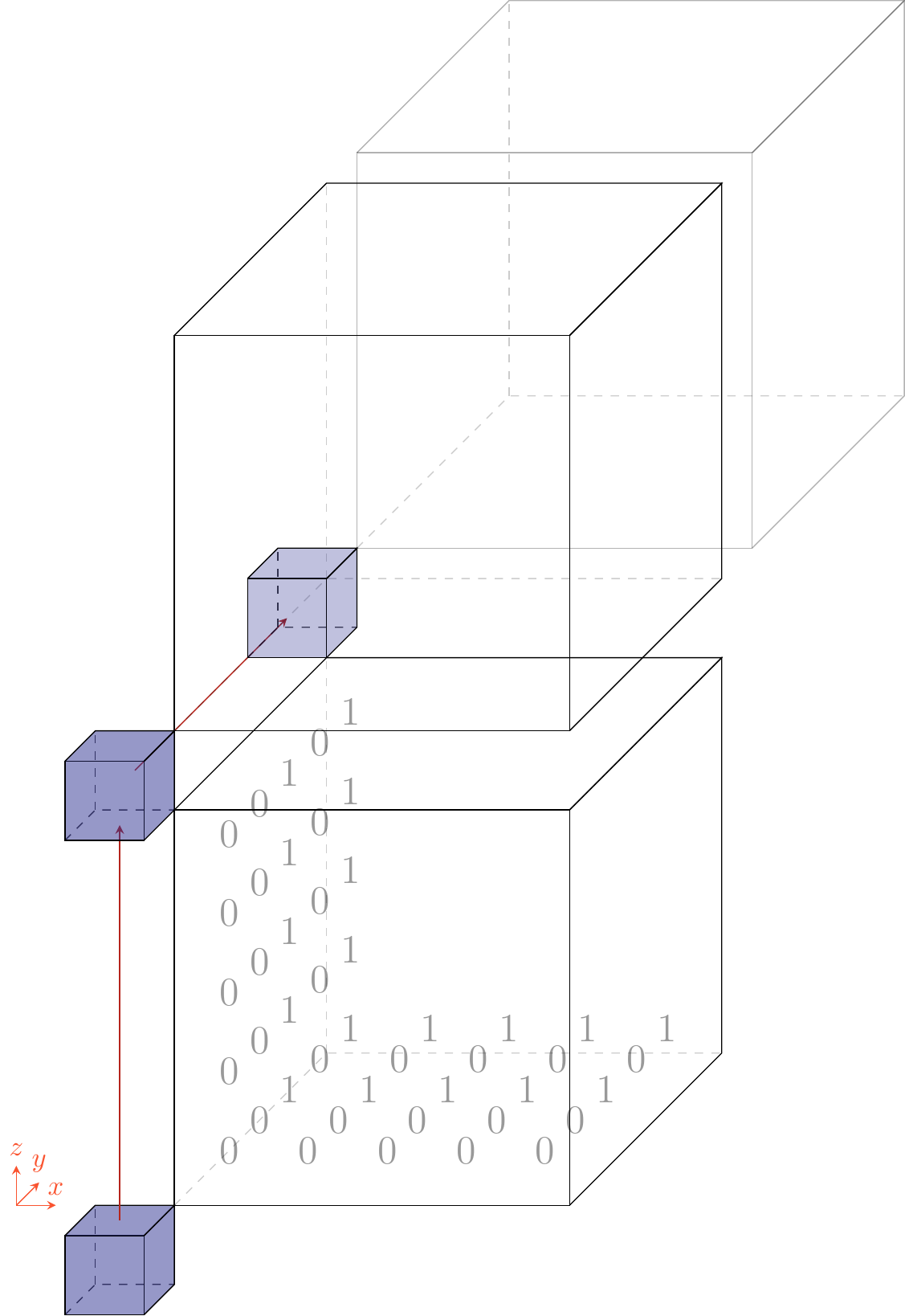}
        \caption{Meta-tiles of large cubes,
            with the adjacency rules represented by the arrows. Myers' tiles are placed in the darker cubes, in the $(yz)$ plane.}
        \label{fig:pentes_1:transitions_norm}
      \end{center}
    \end{minipage}
    \begin{minipage}{.1\textwidth}
        ~
    \end{minipage}%
    \begin{minipage}{.44\textwidth}
      \begin{center}
        ~~~\includegraphics[scale=0.6]{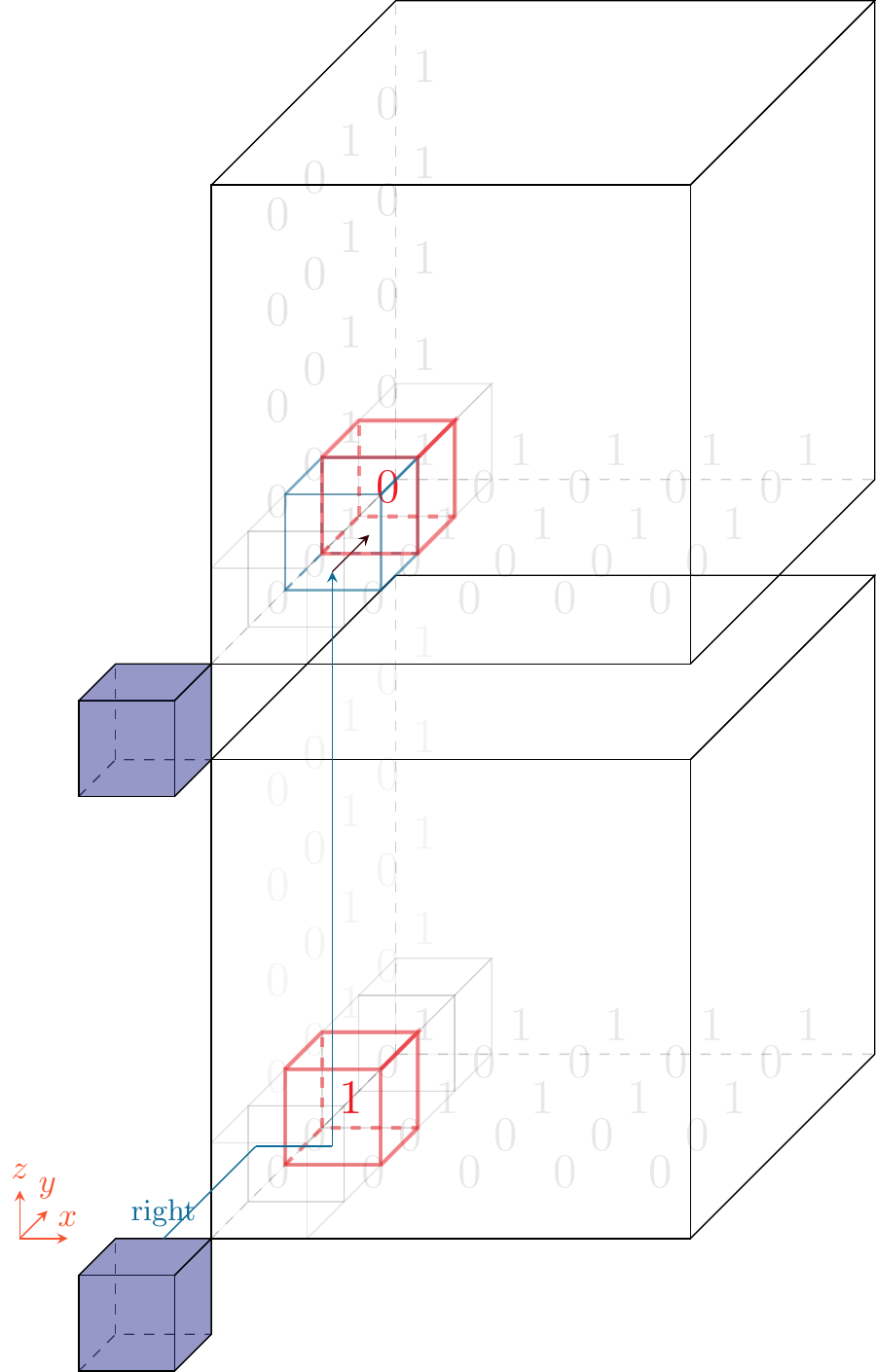}
        \caption{Moving the red cube when the head of $M_O$ moves.}
        \label{fig:pentes_1:marqueur_lecture_move}
      \end{center}
    \end{minipage}
    \end{figure}

  \item[Layer $A$]
  This last layer forces the apparition of 1-periodic configurations. Using two cubes (\includegraphics[width=7pt]{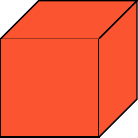} and \includegraphics[width=7pt]{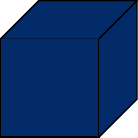}), superimposed only with \includegraphics[width=7pt]{figures/pentes_1/tuiles/blanc.pdf} and borders of big cubes. We impose that blue/red neighbors have the same color.
  It is easy to see that the color is uniform inside a cube and spread to two opposite corners of cubes. Thus all the cubes along $(p,q,r)$ have the same color and there exists at least one 1-periodic configuration.
\end{description}

Now we prove that this construction does what we claim, finishing the proof of Theorem~\ref{th:sigma2}.

\subsection{Every slope \texorpdfstring{$\theta$}{theta} of $X_M$ is accepted by \texorpdfstring{$M$}{M}.}
Let $\theta = (\frac{p}{r}, \frac{q}{r})$ be a slope, by construction every periodic configuration along this slope is formed with cubes of the same size $p$, shifted with the same offset $(q,r)$.
Every cube has the same content, which corresponds to an execution of $M$.
Cubes being of finite size, every execution is a halting execution of $M$.
Let's take $(p,q,r) = 2^k(p',q',r')$, with $p',q',r'$ odds.
Thanks to the layer P, the input of $M$ is $(p',q',r')$, then $M$ accepts $(\frac{p'}{r'}, \frac{q'}{r'}) = (\frac{p}{r}, \frac{q}{r}) = \theta$.

\subsection{Accepting inputs of \texorpdfstring{$M$}{M} are slopes of $X_M$.}
If $M$ accepts the input $(p,q,r)$, there exists a time $t$ and a space $a$ on the working tape, $b$ on the oracle tape, in which the machine $M$ halts.
Then, the cube of size $m = 2^{\lceil \log t\rceil } p \geq a,b,t$ can contain the computation of $M$.
The configuration formed by cubes of size $m$ and of offset $(n,o) = 2^{\lceil \log t\rceil }(q,r)$ is of slope $(\frac{m}{o}, \frac{n}{o}) = (\frac{p}{r}, \frac{q}{r})$.
\end{proof}

\section{Open Problems}
The problem of deciding if all configurations of a 2D SFT are aperiodic is well-known to be \pizu. Proving the other direction of the conjecture would require the study of a very similar problem: deciding if there \textbf{exists} a periodic configuration in a given SFT.
Four our purpose, one needs to prove that the problem of the existence of an aperiodic configuration is \pizu or \sizd.
However, we aren't aware of any study of this problem, not even a simpler bound like \pizd.
Our quick look at it suggests that this could be a very challenging problem to tackle.
Yet, it seems interesting by itself, as it would likely lead to a better understanding
of periodicity and aperiodicity in SFTs.

\section*{Acknowledgements}
The authors would like to thank anonymous reviewers who pointed
out a mistake in a previous version of the paper.

\smallskip

\noindent
This work was supported by grant TARMAC ANR 12 BS02 007 01.

%
%
\bibliographystyle{abbrv}
\bibliography{biblio,biblio-pv,publis}

\end{document}